\documentclass[11pt,twoside]{article}

\usepackage{asp2006}
\usepackage{epsf}
\usepackage{graphicx}
\usepackage{lscape}
\usepackage{cite}
\usepackage{ulem}

\begin{document}
\title{A closer look at the flaring feature in the M\,87 jet}  

\author{C.~S.~Chang\altaffilmark{a,1}, E.~Ros\altaffilmark{b,a}, Y.~Y.~Kovalev\altaffilmark{c,a}, and M.~L.~Lister\altaffilmark{d}}
\affil{\small{\altaffilmark{a}Max-Planck-Institut f\"ur Radioastronomie, Auf dem H\"ugel 69, D-53121 Bonn, Germany}}
\affil{\small{\altaffilmark{b}Departament d'Astronomia i Astrof\'{\i}sica, Universitat de Val\`encia, E-46100 Burjassot, Spain}}
\affil{\small{\altaffilmark{c}Astro Space Centre of Lebedev Physical Institute, Profsoyuznaya 84/32, 117997 Moscow, Russia}}
\affil{\small{\altaffilmark{d}Department of Physics, Purdue University, 525 Northwestern Avenue, West Lafayette, IN 47907, USA}}

\altaffiltext{1}{\scriptsize{Member of the IMPRS of Astronomy and Astrophysics; member of the ESTRELA network.}}
\altaffiltext{2}{\scriptsize{Data are from the 2\,cm\,Survey (Kellermann et al. 2004) and MOJAVE (Lister et al. 2009) programs.}}

\begin{abstract} 
The radio-loud active galactic nucleus in M\,87 hosts a powerful jet fueled by a super-massive black hole in its center. A bright feature 80\,pc away from the M\,87 core has been reported to show superluminal motions, and possibly to be connected with a TeV flare observed around 2005. To complement these studies and to understand the nature of this feature, we analyzed 2\,cm VLBI data from 15 observing runs between 2000 and 2009. This feature is successfully detected at the milli-Jansky level from 2003 to 2007. Our detections show that its milli-arcsecond structure appears to be extended with a steep spectrum, and no compact or rapidly moving features are observed. Our results do not favor a blazar scenario for this feature. 
\end{abstract}

\paragraph{Background}
The active galaxy M\,87 hosts a very powerful one-sided jet ejected from its nucleus, which is believed to have a super massive black hole (SMBH); (Harms et al. 1994) in the center. \textit{Hubble Space Telescope} (\textit{HST}) observations
in 1999 showed a bright knot (HST-1) in the jet, located 80\,pc away from the core (Biretta et al.\,1999), and displaying superluminal motion up to 6\,$c$. This knot is  active in radio,
optical, and X-ray bands. VLBA 20\,cm observations show that HST-1 has sub-structures and superluminal components with speeds of up to 4\,$c$ (Cheung et al. 2007). High-energy observations suggest that HST-1 could
be the origin of the TeV emission detected in M\,87 in 2005 by the HESS telescope (Aharonian et al. 2006).
Comparing with the results in the near ultraviolet (Madrid 2009), \textit{Chandra} soft X-rays (Harris et al. 2006), and
VLA 2\,cm observations (Cheung et al. 2007), the
light curves of HST-1 reach a maximum in 2005, while the resolved
core shows no correlation with the TeV flare. Therefore, the TeV
emission from M\,87 was proposed to originate in HST-1 (Harris et al. 2008), and led the authors to suggest that HST-1 has a blazar nature.
It is widely accepted that AGN blazar properties arise from the region near to the SMBH; however, HST-1 is 80\,pc away from the core. Higher resolution observations can reveal additional information on this question.


\begin{figure*}[t]
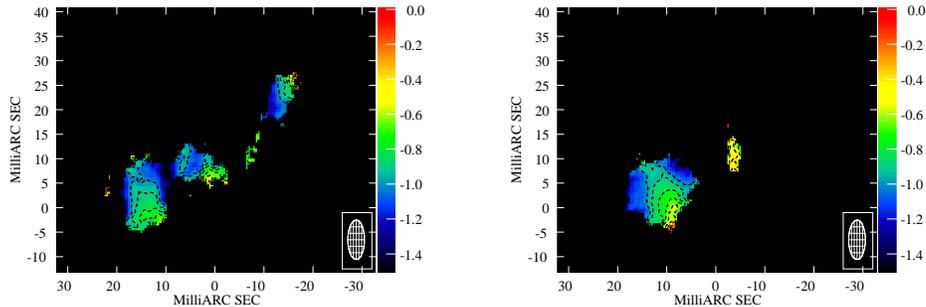

 \centering
 \begin{minipage}[!b]{.49\textwidth}
   \centering
   \includegraphics[width=0.9\textwidth]{changcs_f1.ps}
 \end{minipage}
 \begin{minipage}[!b]{.49\textwidth}
  \centering
  \includegraphics[width=0.9\textwidth]{changcs_f2.ps}
 \end{minipage}
 \begin{minipage}[t]{1\textwidth}
   \caption{\small{Spectral index maps of HST-1 obtained with VLBA 2\,cm and 20\,cm observations at 2005.3 (left) and 2005.8 (right). The restoring beam has a size of 8$\times$3.4\,mas at P.A.\,=\,0$^{\circ}$. }}
 \label{fig:SI_map}
 \end{minipage}

\end{figure*}


\paragraph{Results}
We analyzed 15 VLBI 15\,GHz observing runs of M\,87 from 2000 to 2009$^{\mathrm{2}}$. By applying VLBI wide-field imaging techniques, we could image HST-1 with high resolution. The typical resulting beam size is about 2$\times$1\,mas by applying tapering of Gaussian 0.3 at 200\,M$\lambda$. In order to recover better the extended structure, we downgraded the resolution to a larger beam size (8$\times$3.4 mas at P.A. 0$^{\circ}$).

We detected HST-1 at six epochs from 2003 to early 2007. HST-1 has a total flux density of 3\,mJy to 23\,mJy; the peak surface brightness varies from 1\,mJy\,beam$^{-1}$ to 4\,mJy\,beam$^{-1}$ (high resolution) and 2\,mJy\,beam$^{-1}$ to 10\,mJy\,beam$^{-1}$ (downgraded resolution). We derive the brightness temperatures, T$_{\mathrm{b}}$, of the M\,87 core and HST-1 to be no higher than 5$\times$10$^{9}$\,K and 9$\times$10$^{6}$\,K, respectively. By tracing the peak component of HST-1, we derive a range for the apparent projected speed from 0.55$\pm$0.03\,$c$ to 0.82$\pm$0.03\,$c$. Comparing 20\,cm observations in 2005 and our results, we derive an average spectral index over HST-1 which is steep, about $-$0.8. Figure \ref{fig:SI_map} illustrates the spectral index maps of this region. To conclude, we find HST-1 to be extended over 2-3\,pc (50\,mas), and no compact feature is observed in this region. We do not favor the interpretation of HST-1 having a blazar nature. It is still not clear if HST-1 could be the origin of the TeV emission from M\,87; however, our results do not support this view.


\begin{thebibliography}{}
\bibitem[Aharonian et al.(2006)]{aharonian06}
Aharonian, F., et al. 2006, Science, 314, 1424
\bibitem[Biretta et al.(1999)]{biretta99}
Biretta, J. A., et al. 1999, ApJ, 520, 621
\bibitem[Cheung et al.(2007)]{cheung07}
Cheung, C. C., Harris, D. E., \& Stawarz, \L. 2007, ApJ, 663, L65
\bibitem[Harms et al.(1994)]{harms94}
Harms, R. J., et al. 1994, ApJ, 435, L35
\bibitem[Harris et al.(2006)]{harris06}
Harris, D. E., et al. 2006, ApJ, 640, 211
\bibitem[Harris et al.(2008)]{harris08}
Harris, D. E., et al. 2008, in ASP Conference Series, Vol. 386, 80
\bibitem[Kellermann et al.2004]{kellermann04}
Kellermann, K. I., et al. 2004, ApJ, 609, 539
\bibitem[Lister et al.(2009)]{lister09}
Lister, M. L., et al. 2009, AJ, 137, 3718
\bibitem[Madrid (2009)]{madrid09}
Madrid, J. P. 2009, AJ, 137, 3864


\end{thebibliography}
\end{document}